\documentclass[twocolumn,article]{aastex701}
\usepackage{xcolor}
\usepackage{amsmath}
\usepackage{lipsum}
\usepackage{comment}
\usepackage{makecell}
\usepackage{csvsimple}
\usepackage{xspace}
\usepackage{longtable,array,booktabs}
\usepackage{hyperref}

\newcommand{\msun}{\,\mathrm{M}_\odot}

\defcitealias{2026ApJ...996...69M}{M26}
\newcommand{\miller}{\citetalias{2026ApJ...996...69M}\xspace}

\defcitealias{2024A&A...686A..42H}{H\&R24}
\newcommand{\hunt}{\citetalias{2024A&A...686A..42H}\xspace}

\accepted{July 14, 2026}

\begin{document}

\title{Investigating the Observational Progenitor Mass Gap in the White Dwarf Initial-Final Mass Relation. I. Cluster Census and Characterization of the First White Dwarfs in the Gap}

\correspondingauthor{David R. Miller}
\email{drmiller@phas.ubc.ca}

\author[0000-0002-4591-1903]{David R. Miller}
\affiliation{Department of Physics and Astronomy, University of British Columbia, Vancouver, BC V6T 1Z1, Canada}
\email{drmiller@phas.ubc.ca}

\author[0000-0001-9739-367X]{Jeremy Heyl}
\affiliation{Department of Physics and Astronomy, University of British Columbia, Vancouver, BC V6T 1Z1, Canada}
\email{heyl@phas.ubc.ca}

\author[0000-0001-9873-0121]{Pier-Emmanuel Tremblay}
\affiliation{Department of Physics, University of Warwick, Coventry, CV4 7AL, UK}
\email{P.Tremblay@warwick.ac.uk}

%%%%%%%%%%%%%%%%%%%%%%%%%%%%%%%%%%%%%%%%%%%%%%%%%%%%%%%%%%%%%
%%%%%%%%%%%%%%%%%%%%%%%%%%%%%%%%%%%%%%%%%%%%%%%%%%%%%%%%%%%%%

\begin{abstract}

In recent years, Gaia has been the primary driver of the expansion of the white dwarf (WD) initial-final mass relation (IFMR) in open clusters. The increased sample size has highlighted a pronounced observational gap at progenitor masses of $\simeq2$--$2.7\msun$, with no spectroscopically confirmed cluster-member WDs in this range. Analysis of the Milky Way open cluster census shows that this absence is primarily driven by the scarcity of appropriately aged, nearby clusters capable of hosting Gaia-detectable WDs, implying that deeper, targeted photometry will be required to build a substantial sample in this progenitor mass range. We further searched for previously unexamined Gaia WD candidates in clusters with ages consistent with producing gap progenitors and identified two viable targets, which we observed spectroscopically with Gemini GMOS-N. Both targets yield inferred progenitor masses within the observational gap, making them the first spectroscopically confirmed cluster-member WDs in this progenitor mass range. The NGC~6991 candidate, in particular, is confirmed as a DA WD with an inferred progenitor mass of $2.12^{+0.04}_{-0.15}\msun$, providing further support for a non-monotonic trend in the IFMR.

\end{abstract}


%%%%%%%%%%%%%%%%%%%%%%%%%%%%%%%%%%%%%%%%%%%%%%%%%%%%%%%%%%%%%
%%%%%%%%%%%%%%%%%%%%%%%%%%%%%%%%%%%%%%%%%%%%%%%%%%%%%%%%%%%%%

\section{Introduction}
\label{sec:intro}

The initial-final mass relation (IFMR) connects the birth mass of a star to its white dwarf (WD) remnant. Its form encodes the effects of stellar mass loss, mixing, and envelope evolution, and serves as a key constraint on stellar evolution theory. More broadly, the IFMR informs galactic chemical enrichment and compact object populations, and is critical for population synthesis models used across stellar, galactic, and extragalactic astrophysics.

Open clusters provide the most direct path to constraining the IFMR (e.g., \citealt{2005MNRAS.361.1131F}; \citealt{2009ApJ...692.1013S}; \citealt{2009ApJ...693..355W}; \citealt{2009MNRAS.395.1795C}; \citealt{2009MNRAS.395.2248D}; \citealt{2015ApJ...807...90C}; \citealt{2016ApJ...818...84C}; \citealt{2018ApJ...866...21C}; \citealt{2021ApJ...912..165R}; \citealt{2020NatAs...4.1102M}; \citealt{2022ApJ...926..132H}; \citealt{2022ApJ...926L..24M}; 
\citealt{2026ApJ...996...69M}). Because cluster members are coeval, progenitor masses can be inferred from the WD cooling age and the cluster age using stellar evolution models. The Gaia mission \citep{2016A&A...595A...1G,2018A&A...616A...1G,2021A&A...649A...1G,2023A&A...674A...1G} has transformed these studies by increasing the known WD population by roughly an order of magnitude \citep[see][]{2021MNRAS.508.3877G}, refining membership through precise astrometry, and substantially expanding the census of star clusters \citep[e.g.,][]{2023A&A...673A.114H}.

Building on this foundation, the work of \citet{2026ApJ...996...69M} (hereafter \miller) compiled the largest cluster-based IFMR sample to date, combining new spectroscopy with a uniform reanalysis of literature WDs. The expanded data set more than doubled the Gaia-based spectroscopic sample and increased the total number of high-quality cluster WDs (including non-Gaia sources) by over $50\%$. This large sample highlights a striking feature: an observational gap in the IFMR between progenitor masses of approximately $2$--$2.7\msun$, with no spectroscopically confirmed WDs in this range. While this gap had been present in previous studies  \citep[e.g.,][]{2018ApJ...866...21C,2020NatAs...4.1102M}, the increased sample size of \miller establishes the observational gap as a robust feature. 

The form of the IFMR differs across this gap, with WDs below the gap lying at higher final masses than would be expected from a straightforward extrapolation of the relation above the gap. One possibility is that this discrepancy reflects systematic differences in the cluster populations sampled on either side of the gap. However, \miller found that it persists even when restricting to clusters near the Galactic plane with near solar metallicity, suggesting it is not primarily driven by differing cluster properties. 

One proposed explanation for such an effect is that the IFMR contains a localized non-monotonic feature in this range. \citet{2020NatAs...4.1102M} first proposed a kink near $1.8$--$1.9\msun$, attributed to the onset of carbon-star formation during the thermally pulsing asymptotic giant branch (AGB). Once the surface carbon-oxygen (CO) ratio exceeds unity, carbon-rich dust forms efficiently, couples to radiation pressure, and drives strong mass loss, shortening the AGB phase and curtailing core growth. This mechanism is supported by AGB core mass constraints \citep{2022ApJS..258...43M} and theoretical models with tuned convective overshooting and mass-loss prescriptions \citep{2024ApJ...964...51A}. However, it only appears for a narrow range of overshooting efficiencies. The proposed feature is hinted at in \miller, but the observational gap prevents direct confirmation. Detecting such a feature would constrain convective overshooting, a key uncertainty in predicting stellar mass-loss rates.

In this work, we first investigate whether the absence of cluster WDs with inferred progenitor masses in the $2$--$2.7\msun$ range can be explained by cluster demographics (age, distance, richness, and dispersion). This census analysis is the primary motivation for the study, as it tests whether the gap may reflect the limited population of open clusters in which Gaia could plausibly detect such objects. We then examine the presently available cluster WD sample to identify candidate gap-region objects that have not yet received spectroscopic follow-up, and present the identification and spectroscopic characterization of two such WDs. Finally, we discuss future work needed to substantially expand the sample in this regime. 

We define the observational gap in Section~\ref{sec:lowmassgap}, examine cluster demographics in Section~\ref{sec:cluster_census}, present WD selection, spectroscopy, and IFMR implications in Section~\ref{sec:gapWDs}, and summarize in Section~\ref{sec:summary}.

%%%%%%%%%%%%%%%%%%%%%%%%%%%%%%%%%%%%%%%%%%%%%%%%%%%%%%%%%%%%%
%%%%%%%%%%%%%%%%%%%%%%%%%%%%%%%%%%%%%%%%%%%%%%%%%%%%%%%%%%%%%

\section{Defining the Observational Gap}
\label{sec:lowmassgap}

\begin{figure}[ht]
    \centering
    \includegraphics[width=1.0\columnwidth]{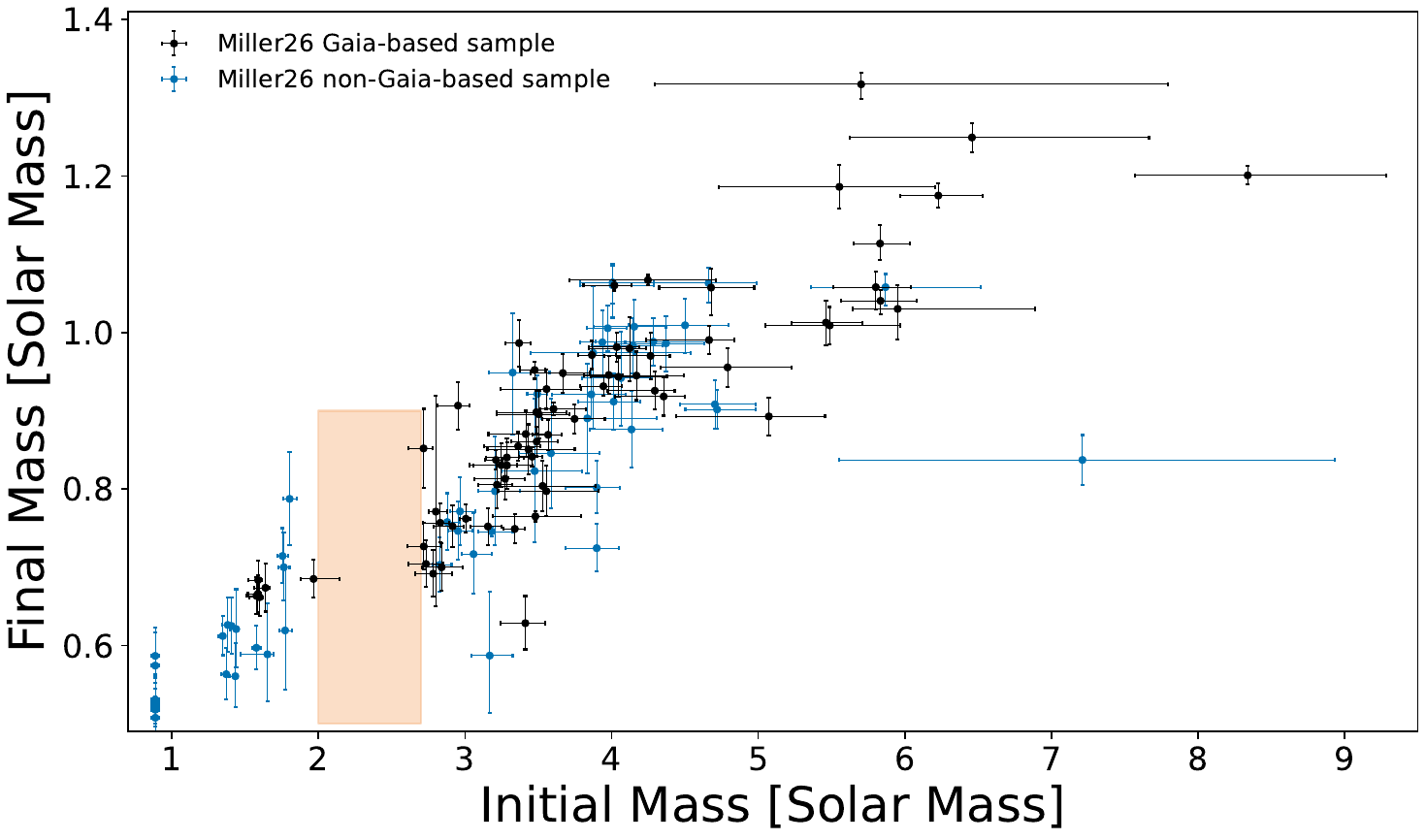}
    \caption{IFMR sample reproduced from \miller. Black points denote WDs with Gaia astrometry supporting membership, while blue points show cluster-member WDs not accessible with Gaia. Shaded orange band marks the $2$--$2.7\msun$ gap.}
    \label{fig:IFMR_initial}
\end{figure}

The observational gap examined here follows that identified in \miller. The nearest best-fit progenitor masses to the gap are at $1.97$ and $2.72\msun$, respectively, but we adopt $2$--$2.7\msun$ as a fiducial definition. Fig.~\ref{fig:IFMR_initial} reproduces the \miller IFMR sample, with the gap highlighted in orange.

To translate the gap into an age range, we consider two complementary definitions. First, we compute progenitor lifetimes for progenitor masses of 2 and $2.7\msun$ using solar metallicity PARSEC isochrones \citep{2012MNRAS.427..127B,2014MNRAS.445.4287T,2014MNRAS.444.2525C,2018MNRAS.476..496F,2022A&A...665A.126N}. Specifically, in each isochrone we identify the zero age main sequence (ZAMS) mass of the lowest-mass star that has reached the onset of the AGB phase, thereby defining a progenitor mass--lifetime relation, which we linearly interpolate at 2 and $2.7\msun$. This yields lifetimes of 1,282 and 596~Myr, respectively. Second, we examine the empirical distribution of clusters with WDs in \miller, which contains no clusters with best-fit ages from 741~Myr (Praesepe) to 1,549~Myr (NGC~752).

Both age ranges are relevant: the first identifies the cluster ages for which newly formed WDs are expected to have progenitor masses in the gap, while the second reflects an age range of clusters without any spectroscopically confirmed WD members. We take the union of these two intervals and define the age gap to span 596--1,549~Myr. For simplicity, we round this to 600--1,550~Myr. This range captures the youngest clusters capable of producing gap-region WDs and includes clusters that could have produced WDs at the upper end of the gap within the last few hundred Myr. We adopt it as a practical target window for identifying clusters likely to host WDs with progenitors in the IFMR gap. 

%%%%%%%%%%%%%%%%%%%%%%%%%%%%%%%%%%%%%%%%%%%%%%%%%%%%%%%%%%%%%
%%%%%%%%%%%%%%%%%%%%%%%%%%%%%%%%%%%%%%%%%%%%%%%%%%%%%%%%%%%%%

\section{Gap-Region Cluster Demographics}
\label{sec:cluster_census}

We begin with the \citet{2024A&A...686A..42H} cluster catalogue (hereafter \hunt), which includes 7,167 Milky Way stellar associations identified in Gaia DR3. Following the authors' recommendations, we restrict to their high-probability subset: clusters that pass the Cluster Significance Test (CST) at $> 5\sigma$ and have a median homogeneity parameter (CMDCl50) $>0.5$. We further exclude globular clusters due to their advanced ages and moving groups, which typically give unreliable stellar parameters, leaving only open clusters. Finally, we retain only clusters with at least 50 member candidates. Less populous clusters are less likely to have formed WDs from gap-region progenitors and often have poorly constrained ages, which would bias the census toward nearby low-population systems. These cuts reduce the sample to 2,454 open clusters.

\subsection{Age--Distance Distribution}

\begin{figure}[ht]
    \centering
    \includegraphics[width=1.0\columnwidth]{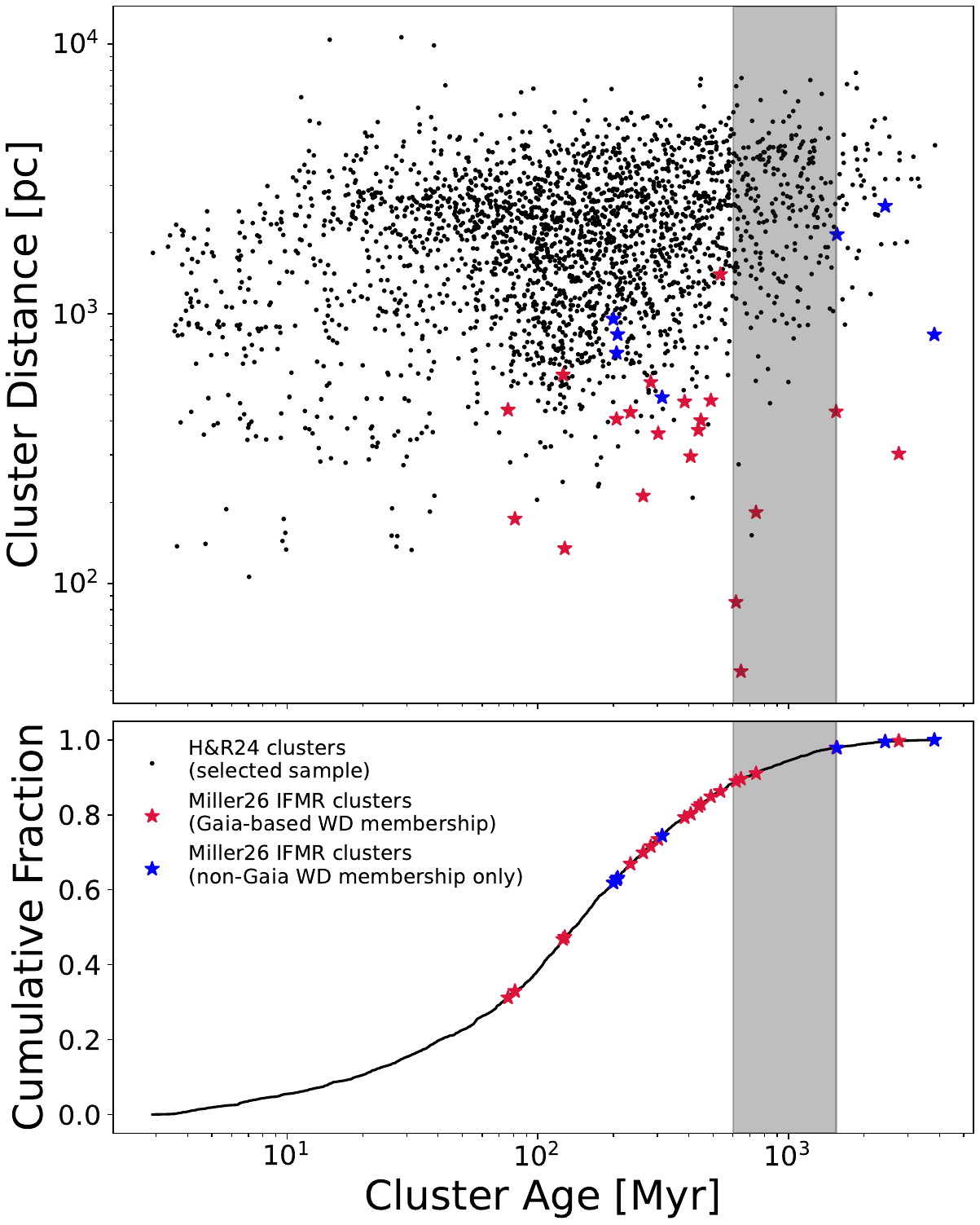}
    \caption{Cluster distance and age distribution for well-populated open clusters in the \hunt catalogue. Red stars mark clusters with at least one Gaia-based WD member in \miller, blue stars mark those with only non-Gaia-based WD associations, with all other clusters shown in black. The grey band indicates the target age range of interest from 600--1,550~Myr.}
    \label{fig:cluster_age_dist}
\end{figure}

Fig.~\ref{fig:cluster_age_dist} shows the age--distance distribution of the selected \hunt clusters (black), with \miller IFMR clusters hosting WDs overplotted (blue/red). For clusters in \miller, we adopt their published ages and the \hunt mean distances; for all other clusters, we use the mean ages and distances from \hunt. Of the clusters in the \miller IFMR sample, we exclude three: the AB~Doradus~Moving~Group and ASCC~47 (not in \hunt), and NGC~6121 (globular cluster).

The cumulative age panel shows a strong skew of the cluster population towards younger ages. While many clusters fall within the relevant age range to have produced WDs from progenitors in the observational gap, relatively few are nearby. Of the eight \miller IFMR clusters in this range, five host at least one Gaia-based WD member: the Hyades, Melotte~111, NGC~752, Praesepe, and Ruprecht~147, the most distant of which lies just 434~pc away. Only three additional clusters within 500~pc are older than 600~Myr (Alessi~3, IC~4756, and OCSN~3). Thus, although many clusters meet the age requirement, very few are both old enough and nearby enough for Gaia to detect gap-region WDs, and most have already been examined in detail.

\begin{figure}[ht]
    \centering
    \includegraphics[width=1.0\columnwidth]{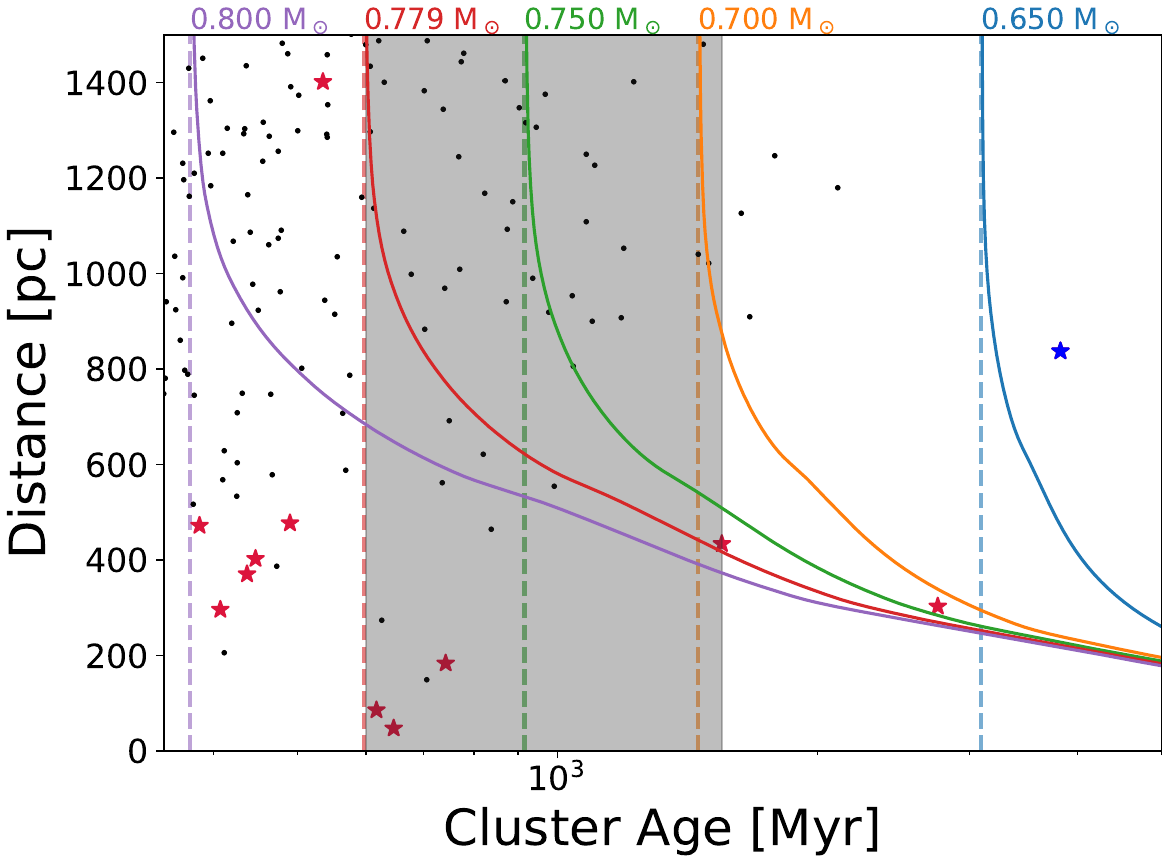}
    \caption{Restricted sample of \hunt clusters shown in age--distance space, focused on the age range of interest. Markers are as in Fig.~\ref{fig:cluster_age_dist}. Colored curves show Gaia G-band detection limits for WD masses of $0.65$--$0.80\msun$ in $0.05\msun$ increments, with an additional curve at $0.779\msun$ chosen because it corresponds to the youngest cluster age that can plausibly produce WDs from progenitors within the gap. Dashed vertical lines indicate the estimated solar-metallicity progenitor lifetime for each WD mass.}
    \label{fig:cluster_age_dmax}
\end{figure}

To remain plausibly detectable, a gap-region WD must (i) have had the time to form, setting a mass-dependent minimum cluster age,  and (ii) remain bright enough to be detected by Gaia, which depends both on the cluster distance and the WD cooling age. We estimate Gaia detection limits by using the monotonic IFMR Fit~2 from \miller to map a grid of WD masses to progenitor masses, converting these to progenitor lifetimes with solar-metallicity PARSEC isochrones, and combining them with the WD cooling sequences of \citet{2020ApJ...901...93B} to compute Gaia $G$ magnitudes as a function of cooling age. This yields detection curves in age--distance space giving the maximum distance at which a WD remains detectable by Gaia. We adopt a nominal limit of $G=20.7$, though the practical limit is typically brighter due to crowding and other observational effects. Fig.~\ref{fig:cluster_age_dmax} shows the resulting curves for WD masses of $0.65$--$0.80\msun$, with vertical dashed lines marking progenitor lifetimes, i.e., the minimum cluster age at which a WD of that mass could be present.

Although the IFMR across the gap is uncertain, this mainly affects inferred progenitor lifetimes used to establish formation-age thresholds. The WD cooling curves themselves are unaffected, and moderate shifts in the progenitor lifetime do not substantially alter which clusters fall within Gaia's detection window. At the low mass end, no clusters lie within Gaia's detection range for gap-region WDs, expect in the case of extremely young objects. Near the young edge of the target age range, more clusters enter the window, but most lie close to the detection limit and several already host known WD members. The sharp spikes at zero cooling age demonstrates that a newly formed WD can be detectable at substantially greater distances than an older WD of the same mass. This is illustrated by NGC~2099, whose Gaia-supported member WD is extremely hot ($T_\textrm{eff}=76{,}000$~K) and therefore detectable at a much larger distance. At higher masses, the nominal detection window includes more clusters, but most are too young to have plausibly formed gap-region WDs.

Aside from very young WDs near zero cooling age, nearly all clusters in the relevant age--distance window already host known Gaia WDs. Several of these WDs have progenitors outside the gap and remain detectable only because their host clusters are particularly nearby. Substantially expanding the gap-region WD sample will likely require photometry deeper than Gaia for clusters in the 600--1,550~Myr age range.

\subsection{Richness and Surface Density}

While \hunt provides a practical census of clusters currently available for Gaia-based studies, it is important to also consider the catalogue completeness at ages relevant to the progenitor-mass gap. If the census is increasingly incomplete for older clusters, the true number of clusters that occupy the age--distance window may be larger than implied by the catalogued sample.

Open clusters lose members and surface density over time through stellar evolution, two-body relaxation, tidal stripping, encounters with giant molecular clouds or the Galactic disc, and dynamical friction \citep{2005A&A...441..117L,2007MNRAS.375..604F,2010MNRAS.409..305L,2019MNRAS.486.5879W}. These processes preferentially remove low-mass clusters \citep{2003MNRAS.338..717B}, making older, low-contrast systems increasingly difficult to distinguish from the field.

\begin{figure}[ht]
    \centering
    \includegraphics[width=1.0\columnwidth]{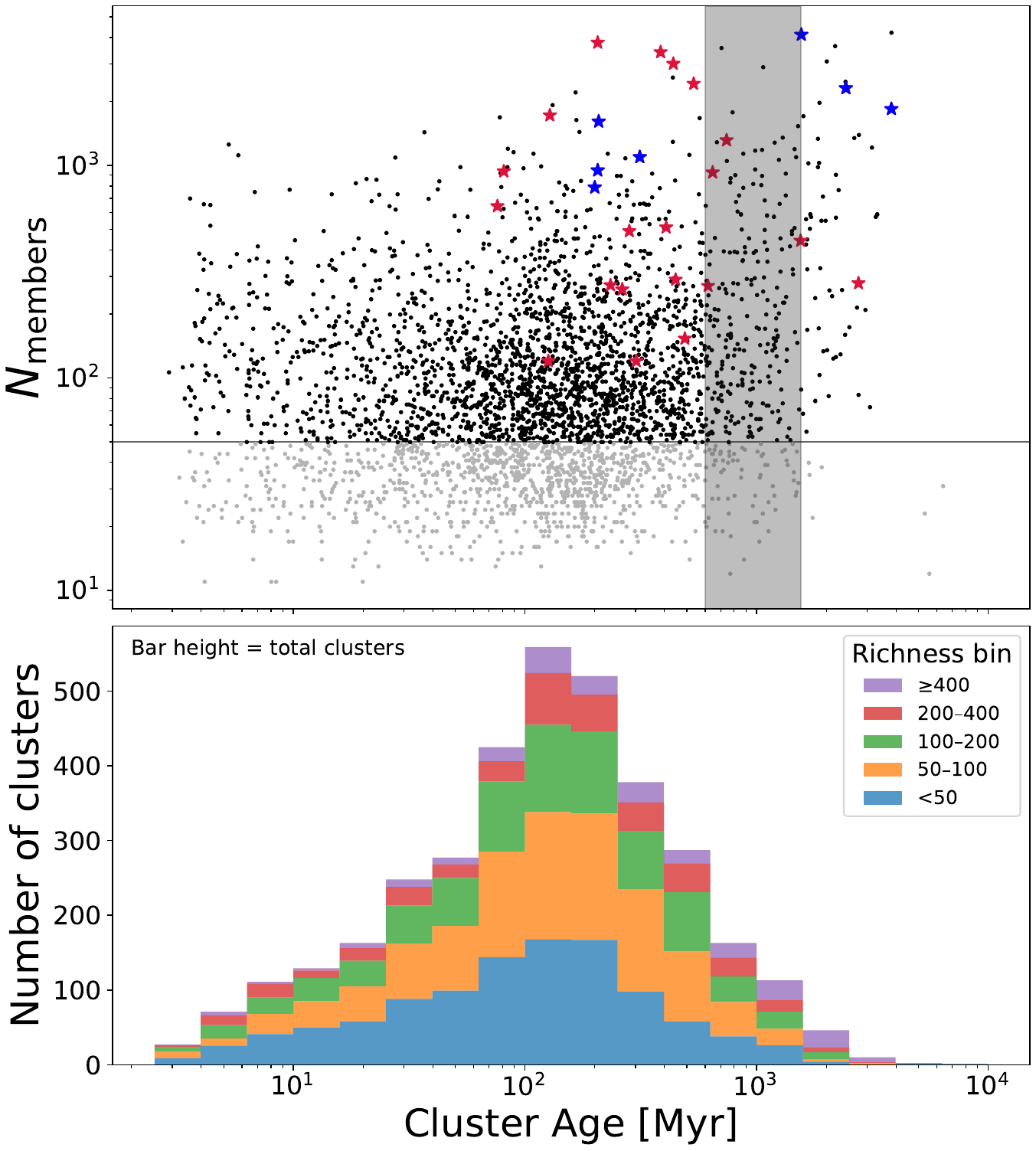}
    \caption{Cluster richness and age distribution for the chosen \hunt sample. The upper panel shows the number of members, with the horizontal line marking $N_{\textrm{members}}=50$, the membership threshold adopted for the primary census analysis. The lower panel shows the age distribution as a stacked, non-cumulative histogram of cluster counts by richness bin. The full height of each bar gives the total number of clusters in that age bin. Markers in the upper panel are as in Fig.~\ref{fig:cluster_age_dist}.}
    \label{fig:cluster_age_richness}
\end{figure}

Fig.~\ref{fig:cluster_age_richness} illustrates this by showing how the membership distribution evolves with age. For this analysis, the membership number restriction is relaxed to include clusters with $N_{\textrm{members}}<50$. The density of clusters with intermediate $N_{\textrm{members}}$ declines significantly toward the oldest ages in the catalogue, while the richest systems increasingly dominate. Since cluster evolution tends to reduce membership over time, this pattern can be interpreted as an age-dependent depletion of low- and intermediate-richness clusters, driven by both physical disruption and the decreasing detectability of low-contrast systems.

This trend is quantified by the stacked, non-cumulative histogram in Fig.~\ref{fig:cluster_age_richness}, which bins the catalogue uniformly in $\log(t/\mathrm{Myr})$ with $\Delta\log t = 0.2$. Near the peak of the census (100--158~Myr), $60.5\%$ of clusters have $N_{\textrm{members}}<100$, while only $4.8\%$ have $N_{\textrm{members}}\geq 400$. In the two bins that most closely bracket the 600--1550~Myr age range of interest, these fractions shift to $23.2\%$ and $16.7\%$, respectively. At older ages ($\geq 1585$~Myr), only $11.9\%$ of clusters have $N_{\textrm{members}}<100$, while $49.1\%$ have $N_{\textrm{members}}\geq 400$. These changing fractions indicate a strongly age-dependent selection function in which sparse and moderately populated clusters are increasingly underrepresented at late times.

\begin{figure}[ht]
    \centering
    \includegraphics[width=1.0\columnwidth]{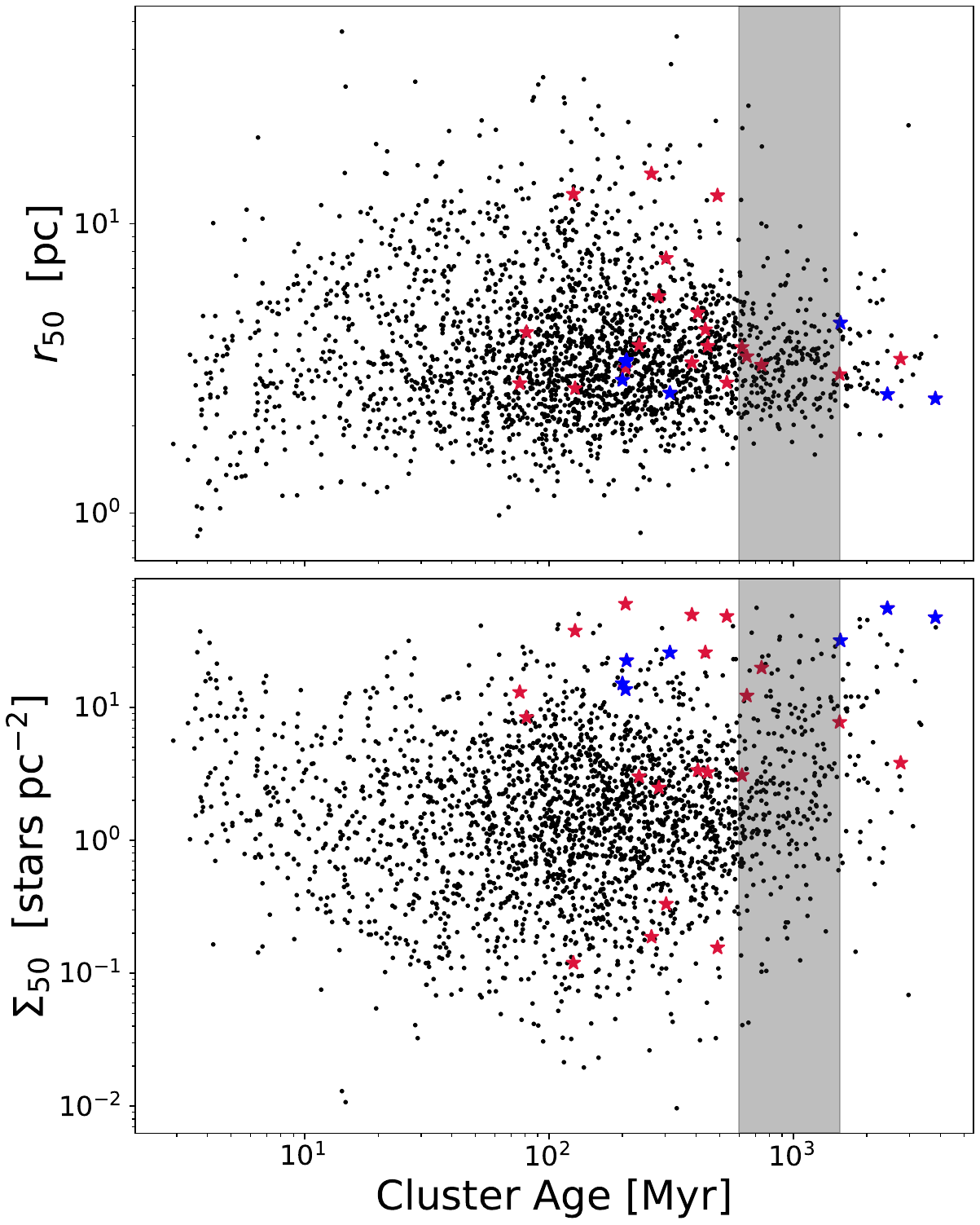}
    \caption{Cluster spatial extent and surface density for the selected sample. The upper panel shows the half-number radius, while the lower panel shows the number-based mean surface density. Markers are as in Fig.~\ref{fig:cluster_age_dist}.}
    \label{fig:cluster_age_r50}
\end{figure}

We also examine cluster dispersal by using the \hunt half-number radius $r_{50}$, defined as the radius containing $50\%$ of members within the tidal radius, and a number-based mean surface density, $\Sigma_{50}=0.5\,N_{\textrm{members}}/(\pi\,r_{50}^{2})$. The distributions of both quantities are shown in Fig.~\ref{fig:cluster_age_r50}. Unlike in the richness analysis, we do not include clusters with $N_{\textrm{members}}<50$, since $r_{50}$ becomes increasingly sensitive to small-number statistics in low-population systems.

Fig.~\ref{fig:cluster_age_r50} shows that particularly compact systems are largely absent at advanced ages, while the lowest-$\Sigma_{50}$ systems are modestly depleted. The oldest catalogue population is therefore biased away from both low-population systems and low-surface-density systems, consistent with physical dispersal and the loss of low-contrast clusters from the recoverable sample.

Taken together, these effects imply that the age--distance window above should be interpreted as a constraint on the catalogue cluster population, but less directly on the underlying population. A substantial population of low richness intermediate-age and older clusters likely remain undiscovered or poorly characterized. Improved astrometry in Gaia DR4 should improve cluster recovery, expanding the census of intermediate-age and older clusters and enlarging the pool of targets for gap-region WD searches. 

%%%%%%%%%%%%%%%%%%%%%%%%%%%%%%%%%%%%%%%%%%%%%%%%%%%%%%%%%%%%%
%%%%%%%%%%%%%%%%%%%%%%%%%%%%%%%%%%%%%%%%%%%%%%%%%%%%%%%%%%%%%

\section{Spectroscopy of Gap-Region White Dwarfs}
\label{sec:gapWDs}

While the previous section shows that few known open clusters occupy the age--distance regime in which Gaia can detect WDs from gap-region progenitors, we can still ask whether any known cluster WD candidates in that regime have not yet been examined spectroscopically for inclusion in Gaia-based IFMR studies.

%%%%%%%%%%%%%%%%%%%%%%%%%%%%%%%%%%%%%%%%%%%%%%%%%%%%%%%%%%%%%

\subsection{Candidate Selection}
\label{sub_sec:wd_candidates}

We begin with the full \hunt cluster catalogue and restrict to high-probability open clusters whose reported $1\sigma$ age range overlaps the 600--1,550~Myr interval of interest, returning 649 clusters. Crossmatching these clusters against the Gaia EDR3 WD catalogue of \citet{2021MNRAS.508.3877G}, and restricting to high-probability WDs ($P_{\textrm WD}>0.75$), returns 52 candidates. We remove candidates in clusters already included in the \miller IFMR sample, leaving nine WDs in six clusters: Alessi~62, IC~4756 (two), NGC~2548, NGC~5822, NGC~6633, and NGC~6991 (three).

\begin{figure*}[ht]
    \centering
    \setlength{\unitlength}{1cm}
    \begin{picture}(13,13.7)(0, 0)
        % Top row
        \put(0.00, 6.53){\includegraphics[width=0.275\textwidth]{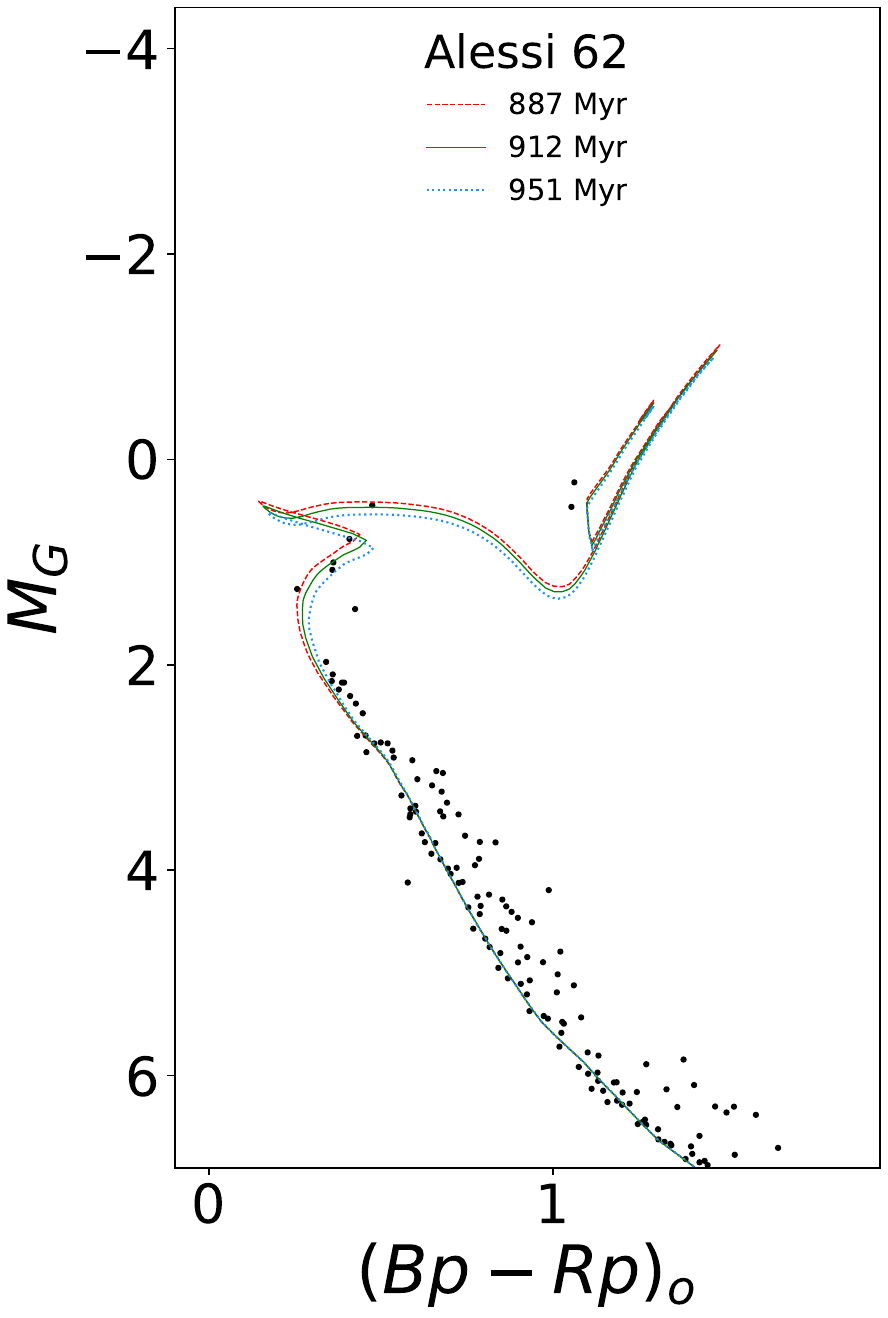}}
        \put(4.91, 6.53){\includegraphics[width=0.223\textwidth]{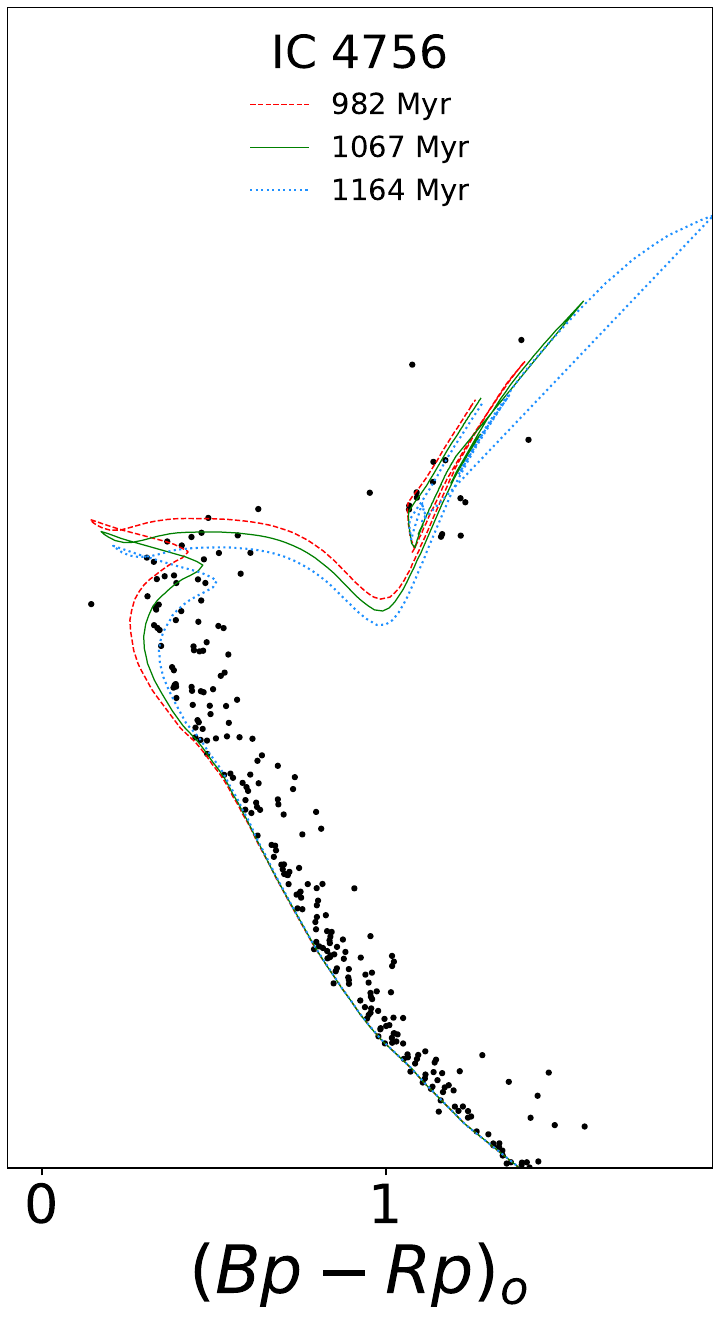}}
        \put(8.89, 6.53){\includegraphics[width=0.223\textwidth]{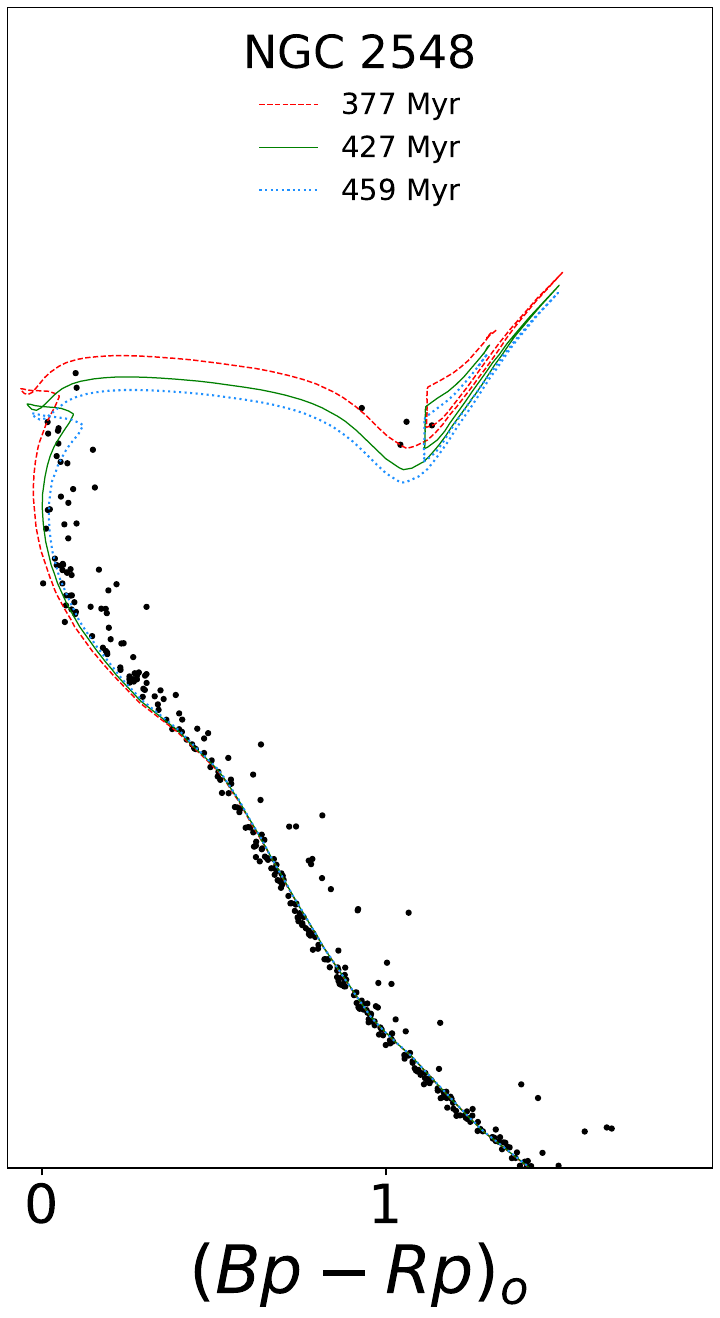}}
        % Bottom row
        \put(0.00, 0.00){\includegraphics[width=0.275\textwidth]{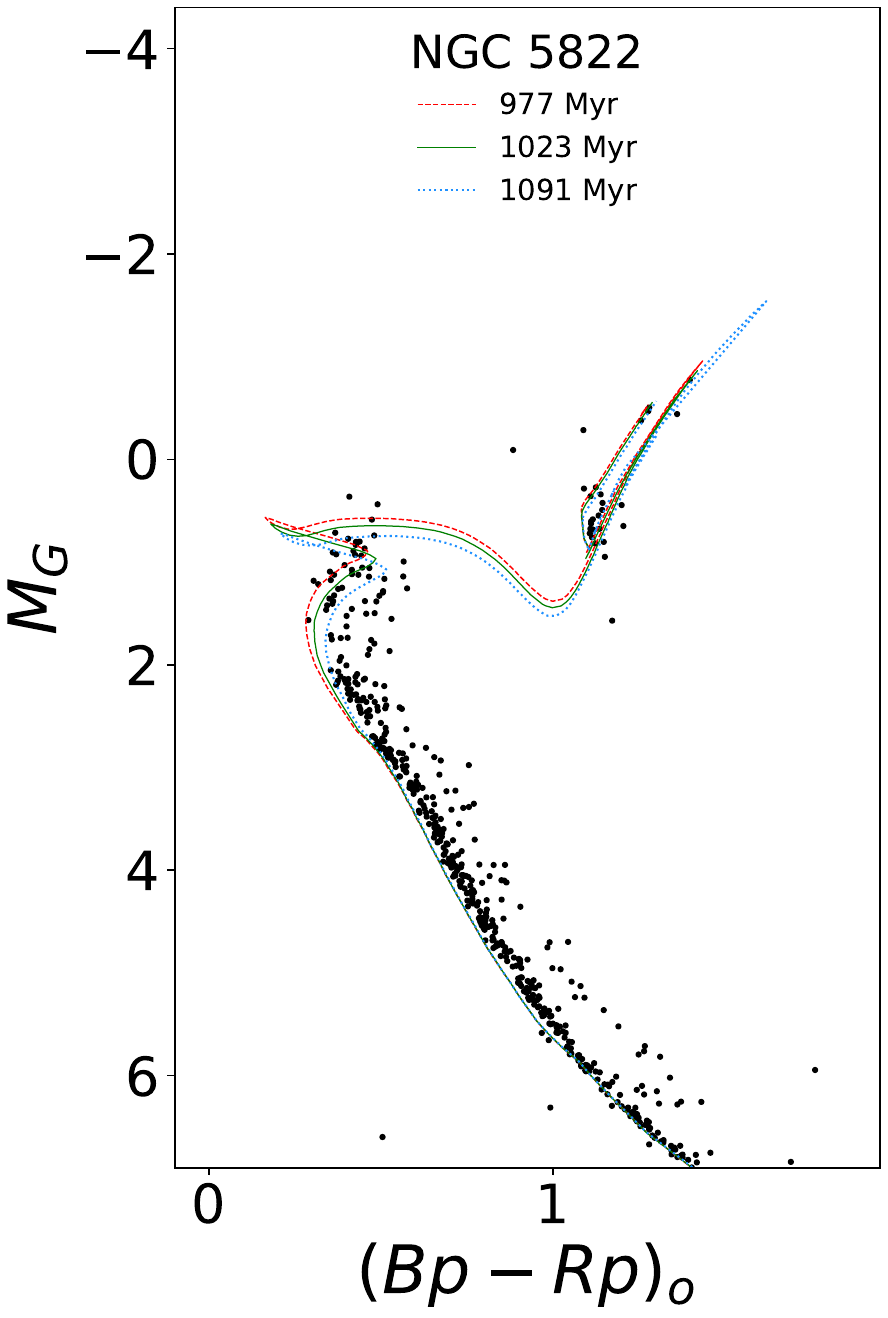}}
        \put(4.91, 0.00){\includegraphics[width=0.223\textwidth]{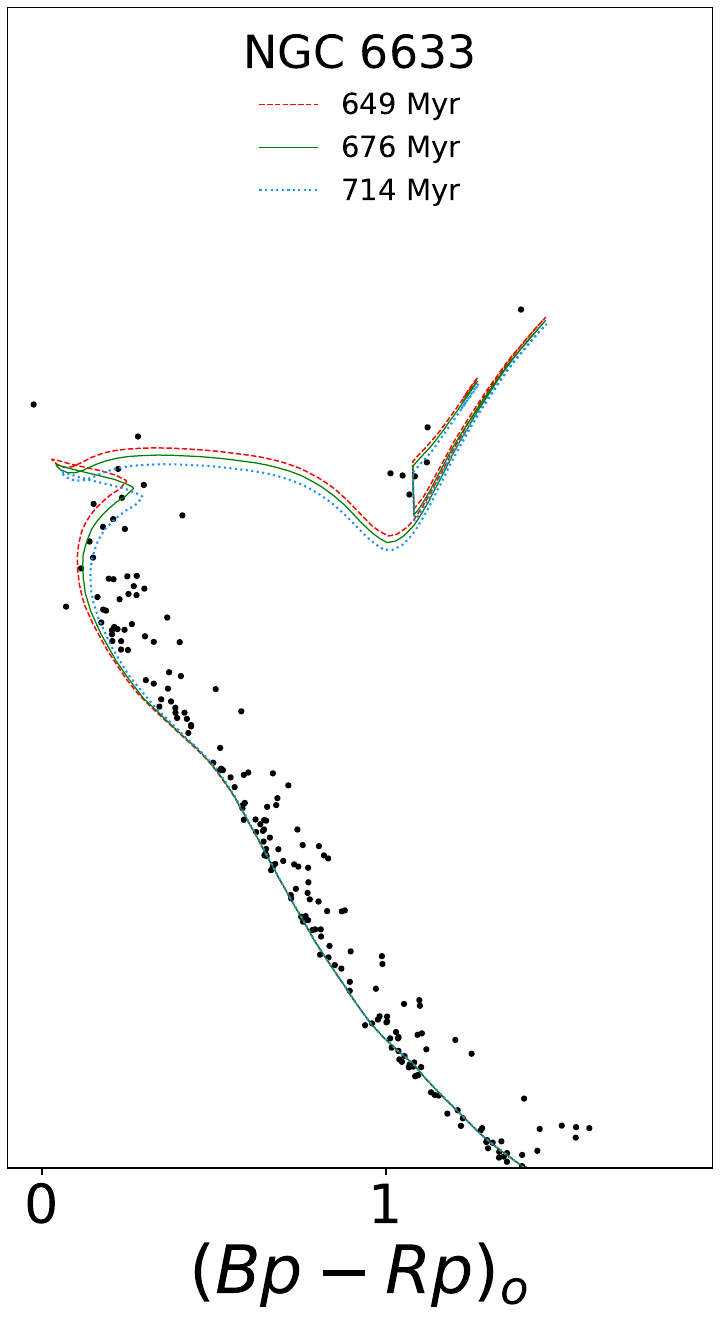}}
        \put(8.89, 0.00){\includegraphics[width=0.223\textwidth]{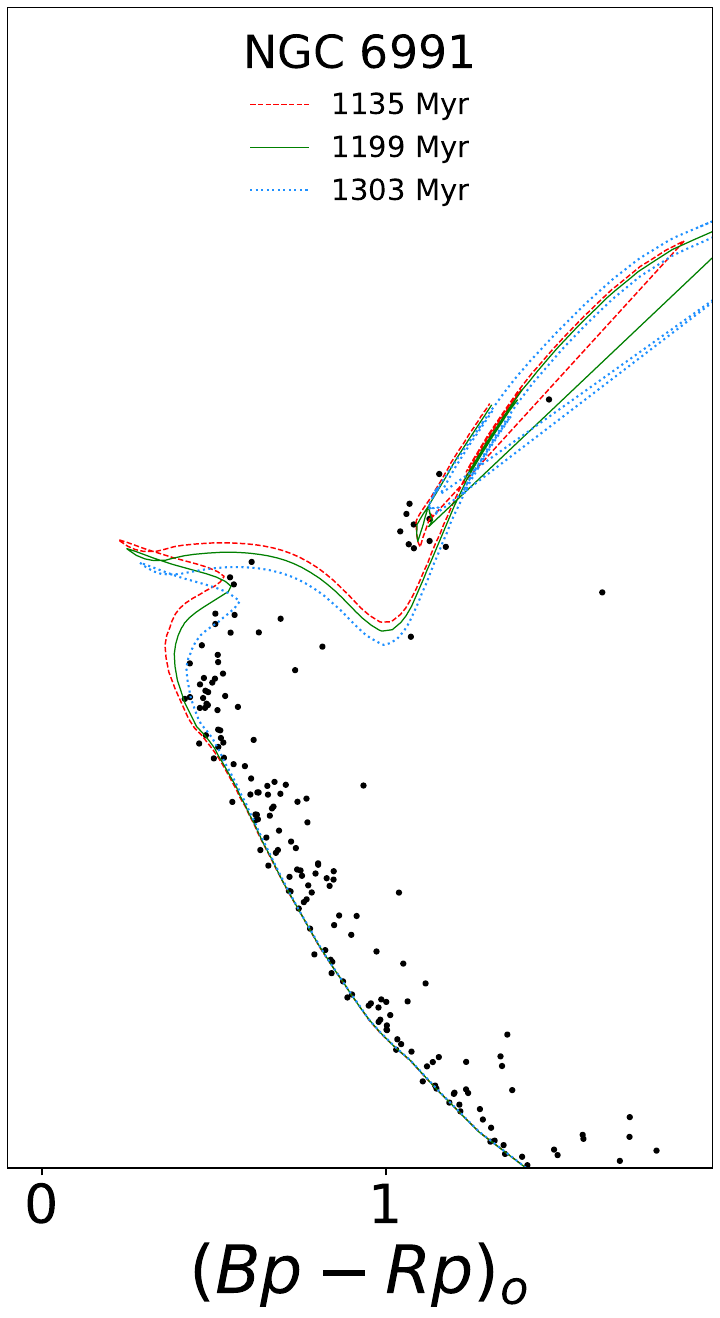}}
    \end{picture}
    \caption{Cluster CMDs for clusters with candidate WD members considered in our isochrone analysis, constructed using stars with $>50\%$ membership probability in \hunt. For each cluster, the best-fit PARSEC isochrone from our heuristic fitting procedure is plotted in green, with the $1\sigma$ age bounds shown in red and blue.}
    \label{fig:cluster_ages}
\end{figure*}

To identify which could plausibly have been born from observational gap progenitors, we derive cluster and WD parameters to test consistency with a single-star origin within the target cluster age window. To determine cluster parameters, we adopt the heuristic MSTO-based PARSEC \citep{2012MNRAS.427..127B,2014MNRAS.445.4287T,2014MNRAS.444.2525C,2018MNRAS.476..496F,2022A&A...665A.126N} isochrone fitting procedure developed by \miller (see their Section~3.3). We perform the fits using \hunt member candidates with membership probabilities $\geq50\%$. In brief, this method combines manual fitting with a semi-automated weighted $\chi^2$ calculation used to define uncertainty bounds. The resulting fits are shown in Fig.~\ref{fig:cluster_ages}, and the adopted cluster parameters are summarized in Table~\ref{tab:cluster_ages_final}. The quoted age uncertainties reflect the internal fitting bounds and do not include potential systematic uncertainties in the stellar models or fitting methodology, for example from unresolved binaries. We find NGC~2548 is too young ($427^{+32}_{-50}$~Myr) to have produced WDs from $2$--$2.7\msun$ progenitors, so we exclude its candidate. The remaining five clusters fall within the target age range.

\begin{table}[ht]
\small
\caption{Properties for clusters hosting WD candidates that potentially lie within the progenitor mass gap.}
\label{tab:cluster_ages_final}
\centering
\begin{tabular}{ccccc}
\toprule
Name &
Distance &
$A_\mathrm{V}$ &
$Z$ &
Age 
\\
 &
[pc] & 
[mag] & 
& 
[Myr]
\\
\midrule
Alessi 62 & 606 & 0.615 & 0.017 & $912^{+39}_{-25}$ \\
IC 4756 & 467 & 0.360 & 0.013 & $1{,}067^{+97}_{-85}$  \\
NGC 2548 & 747 & 0.096 & 0.018 & $427^{+32}_{-50}$ \\
NGC 5822 & 808 & 0.252 & 0.015 & $1{,}023^{+68}_{-46}$ \\
NGC 6633 & 390 & 0.387 & 0.015 & $676^{+38}_{-27}$  \\
NGC 6991 & 557 & 0.238 & 0.015 & $1{,}199^{+104}_{-64}$ \\
\bottomrule
\end{tabular}
\begin{minipage}{\linewidth}
\medskip
{\textbf{Notes}. Distance from \hunt, other parameters from heuristic isochrone fitting.}
\end{minipage}
\end{table}

We estimate WD cooling ages and masses from Gaia photometry by linearly interpolating the \citet{2020ApJ...901...93B}\footnote{\url{http://www.astro.umontreal.ca/~bergeron/CoolingModels}} pure-hydrogen-atmosphere (DA) and CO core model grid, applying extinction corrections based on the cluster reddening derived in this work. We then translate the WD masses to progenitor masses by inverting the \miller IFMR, adopting their full-sample monotonic fit (Fit~2). While this affects the inferred progenitor masses of WDs in the progenitor mass gap, it does not affect our selection, as our aim is to identify candidates plausibly consistent with gap-region progenitors.  We convert progenitor masses to progenitor lifetimes using PARSEC isochrones at the adopted cluster metallicity, and add these lifetimes to the WD cooling ages to obtain a total age estimate. We estimate progenitor masses only for $M_{\textrm WD}\geq0.53\msun$ and do not extrapolate below this limit. This threshold corresponds to the lower-mass boundary of the \miller IFMR. Single-star evolution WDs below this mass would only have been able to form in advanced aged globular clusters and would not have have been born from progenitors within the gap. We summarize the resulting WD parameters along with select Gaia DR3 data in Table~\ref{tab:wd_candidates}.

\begin{table*}
\footnotesize
\centering
\caption{Candidate WD cluster members: select Gaia DR3 data and parameters derived from Gaia photometry.}
\label{tab:wd_candidates}
\begin{tabular}{cccccccccccc}
\toprule
Name & 
Source~ID &
RA & 
Dec & 
Plx &
pmRA &
pmDec &
$G_\textrm{obs}$ & 
$t_\textrm{cool}$ &
$M_\textrm{WD}$ &
$M_\textrm{prog}$ &
$t_\textrm{total}$  \\
      & 
      & 
[deg] & 
[deg] & 
[mas] &
[mas/yr] &
[mas/yr] &
[mag] & 
[Myr] & 
[$\msun$] & 
[$\msun$] & 
[Myr] \\
\midrule
Alessi 62    & 4519349757798439936 & 283.875 &  21.694 & 1.782 &  0.247 & -0.812 & 18.60 &   7 & 0.52 & $\ldots$ & $\ldots$\\
IC 4756 WD1  & 4284010735661963648 & 279.352 &   5.505 & 2.244 &  1.213 & -5.618 & 19.89 & 171 & 0.77 & 3.02 & 593  \\
IC 4756 WD2  & 4283928577215973120 & 279.737 &   5.045 & 2.113 &  1.452 & -5.521 & 17.85 &   4 & 0.33 & $\ldots$ & $\ldots$  \\
NGC 5822     & 5887666586717940224 & 226.164 & -54.407 & 1.038 & -7.093 & -5.507 & 19.47 &   7 & 0.39 & $\ldots$ & $\ldots$  \\
NGC 6633     & 4477214475044842368 & 276.793 &   6.438 & 2.153 &  1.079 & -1.479 & 18.87 &  41 & 0.25 & $\ldots$ & $\ldots$  \\
NGC 6991 WD1 & 2166915179559503232 & 313.505 &  47.803 & 1.959 &  5.414 &  8.791 & 18.90 &  13 & 0.67 & 1.74 & $1{,}809$\\
NGC 6991 WD2 & 2166829112734085248 & 313.419 &  47.296 & 1.725 &  5.333 &  9.140 & 19.85 & 131 & 0.39 & $\ldots$ & $\ldots$  \\
NGC 6991 WD3 & 2163824456685399168 & 314.173 &  46.972 & 1.979 &  5.777 &  8.993 & 20.57 & 480 & 0.49 & $\ldots$ & $\ldots$  \\
\bottomrule
\end{tabular}
\begin{minipage}{\linewidth}
\medskip
{\textbf{Notes}. Candidates are associated with clusters whose \hunt age $\pm1\sigma$ overlap 600--1,550~Myr. Progenitor mass (and thus total age) is not estimated for sources with $M_{\textrm WD}<0.53\msun$.}
\end{minipage}
\end{table*}

Six of the remaining eight candidates have $M_{\textrm WD}<0.53\msun$ (Alessi~62, IC~4756~WD2, NGC~5822, NGC~6633, NGC~6991~WD2, and NGC~6991~WD3), inconsistent with single-star evolution at the corresponding cluster ages. They could be genuine members if binary evolution played a role, but they would not be useful for constraining the IFMR and are therefore discarded. We also consider the additional IC~4756 WD candidate considered by \miller that did not appear in the \citet{2021MNRAS.508.3877G} WD catalogue, but find its CMD position implies a very high mass WD that would not come from a progenitor in the $2$--$2.7\msun$ regime.

This leaves two plausible gap candidates: IC~4756~WD1 and NGC~6991~WD1. Both sources are broadly consistent with their host clusters based on Gaia astrometry, although IC~4756~WD1 has comparatively large astrometric uncertainties, so its membership is less well established from astrometry alone. Spectroscopy provides complementary membership support by testing consistency between spectroscopic WD parameters and the progenitor mass and lifetime implied by the cluster age and IFMR, effectively rejecting outliers.

IC~4756~WD1 yields a slightly younger total age than the cluster, while NGC~6991~WD1 yields a somewhat older age. However, small changes in WD mass can translate into large shifts in progenitor mass and lifetime in this regime, and progenitor mass is additionally sensitive to the adopted IFMR. We therefore retain both candidates as prime follow-up targets. Spectroscopy is required to determine robust WD atmospheric and cooling parameters, enabling a definitive test of whether each source yields an inferred progenitor mass within the $2$--$2.7\msun$ range and a total age consistent with its host cluster.

%%%%%%%%%%%%%%%%%%%%%%%%%%%%%%%%%%%%%%%%%%%%%%%%%%%%%%%%%%%%%

\subsection{Spectroscopic Follow-up}

We obtained spectra for IC~4756~WD1 and NGC~6991~WD1 using the Gemini Multi-Object Spectrograph (GMOS; \citealt{2004PASP..116..425H,2016SPIE.9908E..2SG}) on Gemini-North (GN-2025B-FT-102), with total exposure times of 4,320~s and 1,068~s, respectively. Observations used long-slit mode with a 1" slit and 2x2 binning, with the B480 grating and no filter, centered at 520~nm for approximately 320--720~nm coverage. The spectra were reduced with DRAGONS \citep{2023RNAAS...7..214L}. Atmospheric, cooling, and progenitor parameters are summarized in Table~\ref{tab:wd_params} and described in the following subsections. 

\begin{table*}
\small
\caption{Spectroscopic and derived parameters for WDs followed-up with Gemini GMOS-N.}
\label{tab:wd_params}
\centering
\begin{tabular}{ccccccccc}
\toprule
Name &
Type &
$\log g$ &
$T_\textrm{eff}$ &
$\log(\mathrm{H}/\mathrm{He})$ &
$M_\textrm{WD}$ &
$t_\textrm{cool}$ &
$R$ &
$M_\textrm{prog}$ \\
&
&
$[\mathrm{cm}\,\mathrm{s}^{-2}]$ &
[K] &
&
[$\msun$] &
[Myr] & 
[km] &
[$\msun$] 
\\
\midrule
IC~4756~WD1  & DBA & $8.24\pm0.06$ & $22{,}200\pm600$ & $-4.26\pm0.09$ & $0.75\pm0.04$ & $79^{+14}_{-13}$       & $7{,}600\pm300$ & $2.24^{+0.06}_{-0.08}$ \\
NGC~6991~WD1 & DA  & $8.03\pm0.04$ & $32{,}900\pm200$ & $\ldots$       & $0.67\pm0.02$ & $6.82^{+0.17}_{-0.16}$ & $9{,}100\pm300$ & $2.12^{+0.04}_{-0.15}$ \\
\bottomrule
\end{tabular}
\begin{minipage}{\linewidth}
\medskip
{\textbf{Notes}. For IC~4756~WD1, the spectral fit excluded the 447~nm He line.}
\end{minipage}
\end{table*}

IC~4756~WD1 shows strong He~I and weak H$\alpha$, consistent with a DBA classification. We detect no metal lines or magnetic signatures. NGC~6991~WD1 shows a standard DA spectrum with prominent Balmer absorption and no magnetic field. Both spectra show an absorption feature near 687~nm, which we attribute to telluric $\mathrm{O}_2$; it does not coincide with any significant H or He~I features. 

%%%%%%%%%%%%%%%%%%%%%%%%%%%%%%%%%%%%%%%%%%%%%%%%%%%%%%%%%%%%%%%%

\subsubsection{IC~4756~WD1}

We fit IC~4756~WD1 with the DBA model grid of \citet{2021MNRAS.501.5274C} using an approach similar to \citet{2005ApJS..156...47L}. We begin by fitting the observed spectrum to a synthetic grid along with a polynomial to account for continuum calibration issues. The spectrum is then normalized using fixed continuum points around each line, and the line profiles are simultaneously fit using Levenberg--Marquardt least-squares minimization throughout. Because H abundance affects both Balmer and He~I line strengths, we adopt an iterative fitting procedure following \citet{2011ApJ...737...28B}. We fix $\log(\mathrm{H}/\mathrm{He})$ to fit the He~I lines for the effective temperature ($T_\textrm{eff}$) and surface gravity ($\log g$), then fixing $T_\textrm{eff}$ and $\log g$ to fit H$\alpha$ for $\log(\mathrm{H}/\mathrm{He})$, iterating until convergence within 0.05~dex.

\begin{figure}
    \centering
    \includegraphics[width=1.0\columnwidth]{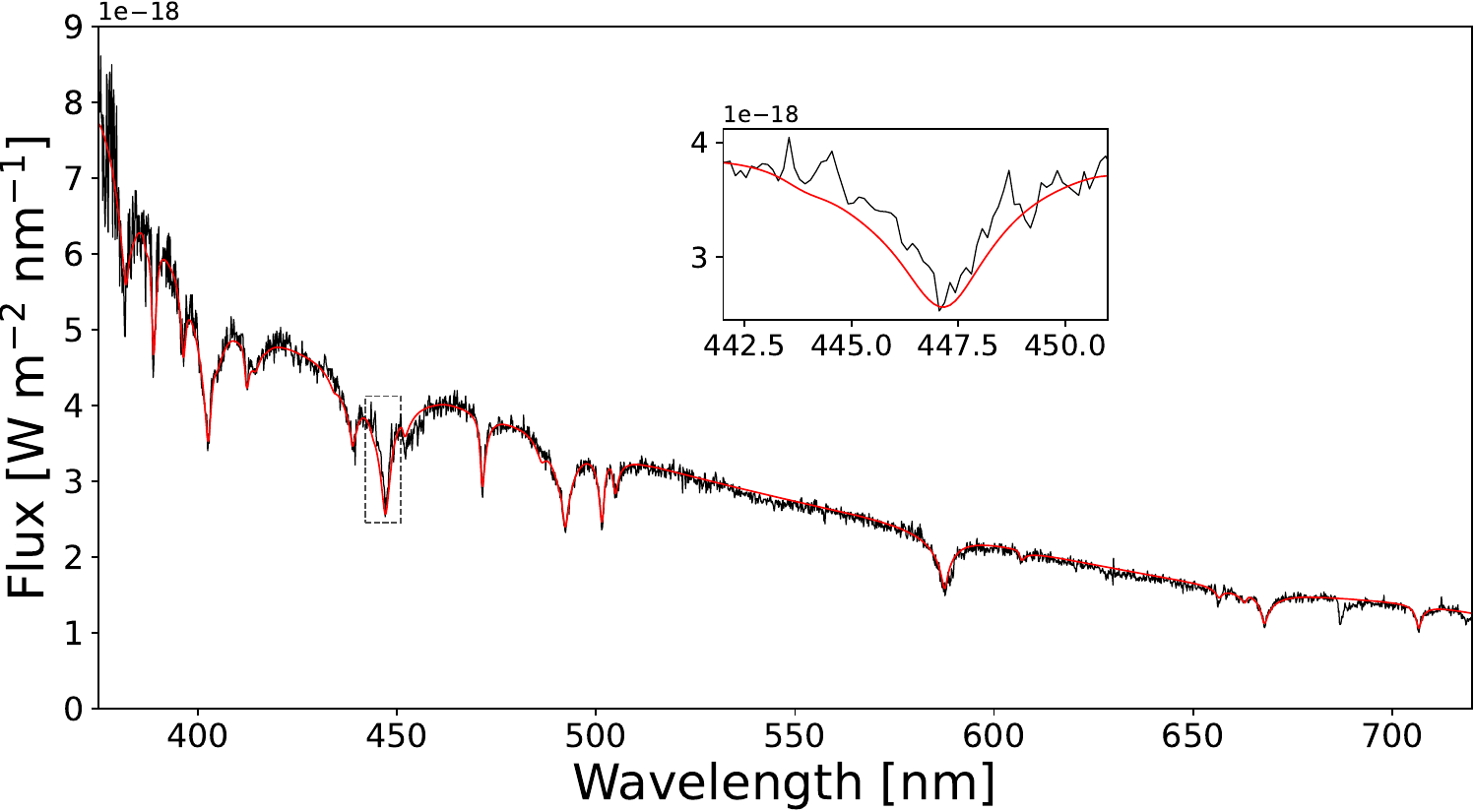} 
    \caption{Gemini GMOS-N spectrum of IC~4756~WD1 with best-fit model from \citet{2021MNRAS.501.5274C} superimposed. Inset highlights the excluded line near $447.1$~nm.}
    \label{fig:IC4756_fit}
\end{figure}

The neutral He~I line near $447.1$~nm prevented convergence. The model profile implied by the best-fit solution to the other lines is significantly broader than the observed line, and the observed line wings show mild asymmetry, suggesting a local normalization issue. This feature falls near a GMOS chip gap; inspection of the individual exposures shows significant exposure-to-exposure variations in the feature near this wavelength, indicating a likely instrumental and/or reduction-related effect rather than a physical line-profile mismatch. Since the simultaneous fit to the remaining strong He~I lines converges cleanly, we exclude this feature from the final fit. The resulting best-fit model is shown in Fig.~\ref{fig:IC4756_fit}.

From $T_\textrm{eff}$ and $\log g$, we estimate the cooling age, mass, and radius of IC~4756~WD1 by interpolating the CO core, thin H envelope pure He atmosphere WD cooling models of \citet{2020ApJ...901...93B}. We then subtract the WD cooling age from the adopted cluster age to obtain the progenitor lifetime, and use PARSEC isochrones at the cluster metallicity to infer the corresponding progenitor mass. The progenitor mass is taken as the lowest-mass star to have reached the onset of the AGB phase, chosen for consistency across the full IFMR sample, as in \miller. The best-fit atmospheric parameters and derived cooling and progenitor properties are reported in Table~\ref{tab:wd_params}.

%%%%%%%%%%%%%%%%%%%%%%%%%%%%%%%%%%%%%%%%%%%%%%%%%%%%%%%%%%%%%%%%

\subsubsection{NGC~6991~WD1}

\begin{figure}
    \centering
    \includegraphics[width=0.7\columnwidth]{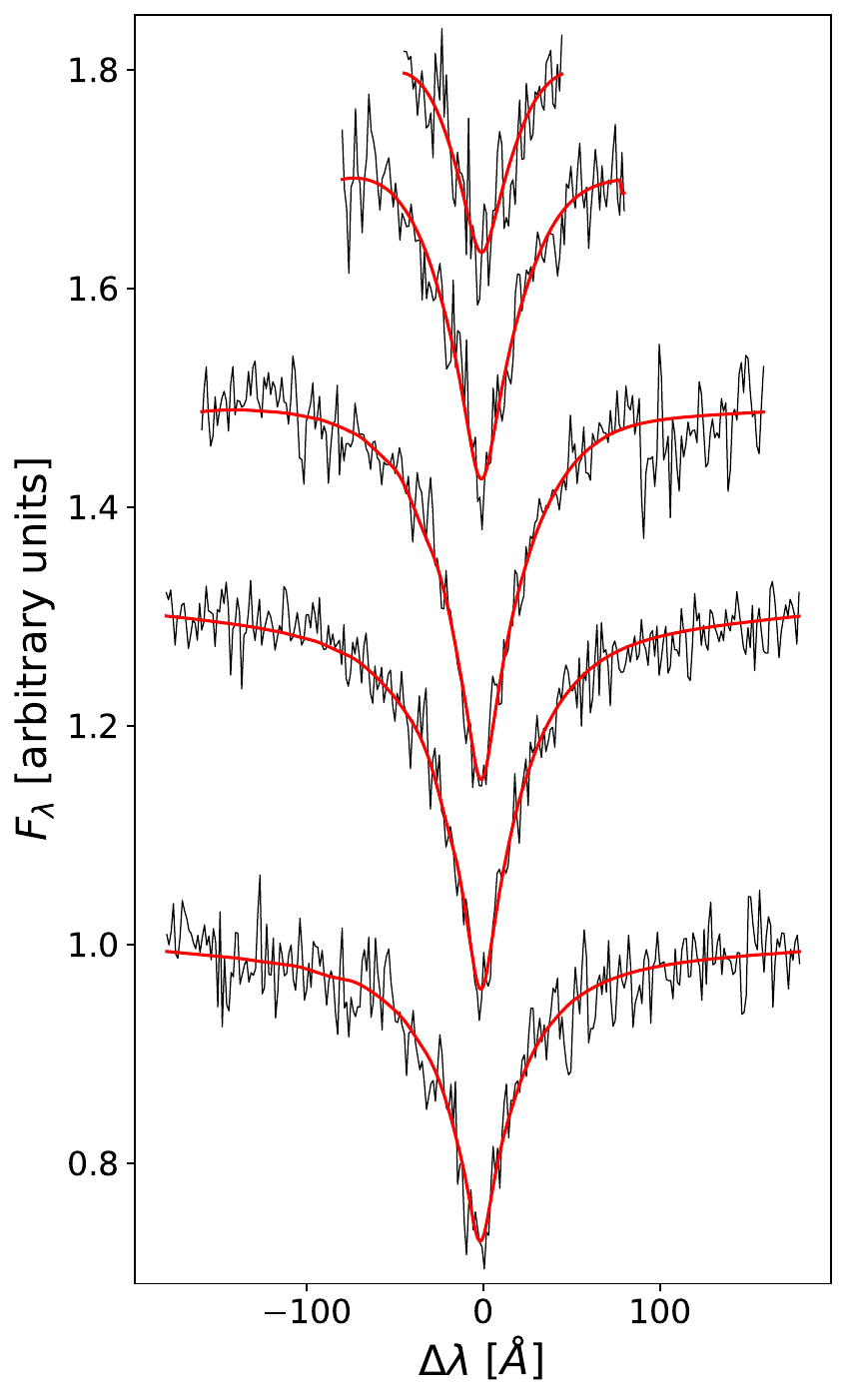} 
    \caption{Balmer line profiles from H$\alpha$ to H$\epsilon$ for NGC~6991~WD1. The best-fitting pure-H model is overplotted in red.}
    \label{fig:NGC6991_fit}
\end{figure}

We fit NGC~6991~WD1 with the non-local thermodynamic equilibrium (NLTE) pure H grid of \citet{2011ApJ...730..128T}. We follow the same fitting technique as above, except that the DA classification requires no abundance ratio and thus no iteration. The Balmer line best-fit model is shown in Fig.~\ref{fig:NGC6991_fit}. We adopt the thick H-layer DA models of \citet{2020ApJ...901...93B} to determine WD cooling parameters and proceed as for IC~4756~WD1 otherwise. The adopted atmospheric fit parameters, along with the inferred cooling and progenitor properties, are summarized in Table~\ref{tab:wd_params}.

%%%%%%%%%%%%%%%%%%%%%%%%%%%%%%%%%%%%%%%%%%%%%%%%%%%%%%%%%%%%%

\subsection{Results and Implications for the IFMR}
\label{sub_sec:ifmr_impact}

\begin{figure*}[ht]
    \centering
    \includegraphics[width=1.0\textwidth]{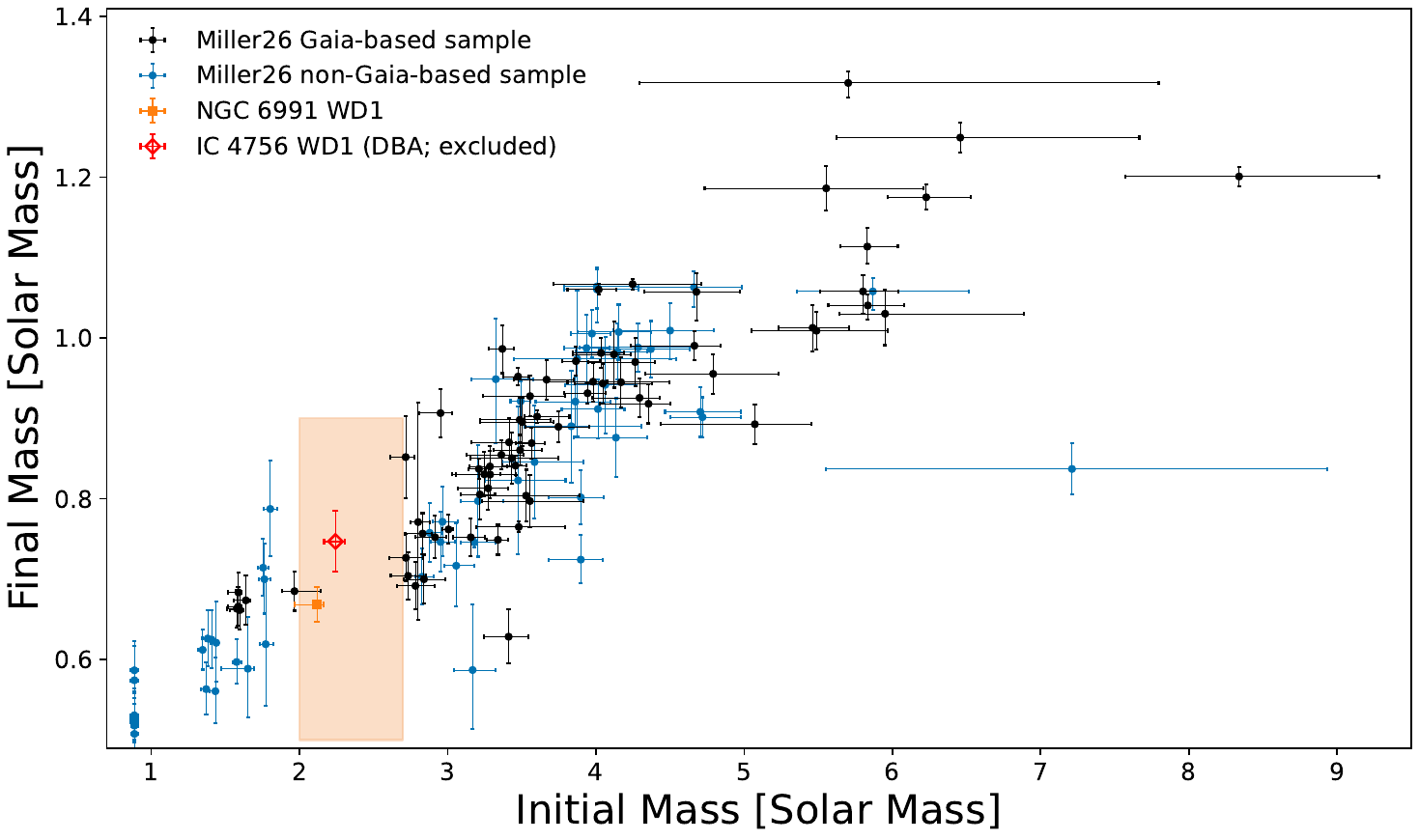}
    \caption{IFMR sample from \miller, as in Fig.~\ref{fig:IFMR_initial}, now including NGC~6991~WD1. The DBA WD IC~4756~WD1 is excluded from the IFMR sample but shown for context.}
    \label{fig:IFMR_final}
\end{figure*}

The fit for NGC~6991~WD1 reveals a young WD with  $t_\textrm{cool}=6.82^{+0.17}_{-0.16}$~Myr and $M_\textrm{WD}=0.67\pm0.02\msun$. With a cluster age of $1{,}199^{+104}_{-64}$~Myr, this implies $M_\textrm{prog}=2.12^{+0.04}_{-0.15}\msun$, within the $2$--$2.7\msun$ gap. Fig.~\ref{fig:IFMR_final} shows the \miller IFMR including this source. Its position within the gap is consistent with the mild non-monotonic trend suggested by \miller, but one additional DA WD is not sufficient to distinguish genuine IFMR structure from systematics in the cluster age, WD parameters, cooling models, or inferred progenitor mass; a much larger sample will be required to better assess the IFMR shape in this regime.

IC~4756~WD1 is a DBA WD with $M_\textrm{WD}=0.82\pm0.03\msun$ and $M_\textrm{prog}=2.29^{+0.07}_{-0.08}\msun$, placing it within the gap and above the DA IFMR trend (Fig.~\ref{fig:IFMR_final}). However, the spectroscopically confirmed cluster DB sample remains small ($\sim6$), and several are magnetic, making it difficult to assess whether this represents a systematic offset for He-atmosphere WDs or an outlier. Although many IFMR studies include DB WDs alongside DAs (e.g., \citealt{2018ApJ...866...21C}; \citealt{2021ApJ...912..165R}), \miller restricted their IFMR sample to DA WDs to reduce systematics. We follow that choice and omit DB WDs, including IC~4756~WD1. While the GMOS spectrum supports a single-star interpretation at the cluster age, the cluster membership of IC~4756~WD1 remains less secure than for NGC~6991~WD1 because its Gaia astrometry is comparatively uncertain. Improved astrometric precision with Gaia DR4 will enable a more stringent membership assessment. The growing cluster DB WD sample warrants future investigation for systematic differences relative to DAs, though substantially more data will be required. 

%%%%%%%%%%%%%%%%%%%%%%%%%%%%%%%%%%%%%%%%%%%%%%%%%%%%%%%%%%%%%
%%%%%%%%%%%%%%%%%%%%%%%%%%%%%%%%%%%%%%%%%%%%%%%%%%%%%%%%%%%%%

\section{Summary and Conclusions}
\label{sec:summary}

This work investigated whether the $2$--$2.7\msun$ IFMR gap highlighted by \miller can be explained by open-cluster demographics, and searched for unexamined Gaia WD candidates in clusters whose ages are consistent with the 600--1{,}550~Myr age window associated with the gap. A census of \hunt clusters shows that few occupy the age--distance window where Gaia can detect gap-region WDs, and most already host cluster member WDs supported by Gaia astrometry, implying that expanding the sample will require deeper photometry. The \hunt census was shown to be increasingly biased against sparse and diffuse systems at older ages, consistent with disruption and decreasing surface density contrast. The age--distance window should therefore be interpreted as primarily a constraint on the catalogue cluster population, not the underlying one. 

We crossmatched \hunt cluster members with the high-probability WD candidates in the Gaia WD catalogue of \citet{2021MNRAS.508.3877G} and identified nine WD candidates in six clusters whose \hunt age uncertainties overlap the target range. Re-deriving cluster parameters following \miller and estimating WD parameters from Gaia photometry reduced the set to two follow-up targets: IC~4756~WD1 and NGC~6991~WD1. Gemini GMOS-N spectroscopy confirms NGC~6991~WD1 as a DA WD with $M_\textrm{prog}=2.12^{+0.04}_{-0.15}\msun$, within the gap and supporting a non-monotonic IFMR trend near the low-mass edge of the gap. IC~4756~WD1 is a DBA WD with  $M_\textrm{prog}=2.29^{+0.07}_{-0.08}\msun$; it lies within the gap but is excluded from the pure DA IFMR. A definitive test of the IFMR across $2$--$2.7\msun$ requires a substantially larger DA WD sample, likely from targeted deep imaging of appropriately aged clusters whose WDs are expected to lie beyond Gaia's practical detection limit.   

%%%%%%%%%%%%%%%%%%%%%%%%%%%%%%%%%%%%%%%%%%%%%%%%%%%%%%%%%%%%%
%%%%%%%%%%%%%%%%%%%%%%%%%%%%%%%%%%%%%%%%%%%%%%%%%%%%%%%%%%%%%

\begin{acknowledgments}
The authors thank Ilaria Caiazzo for helpful comments on the manuscript. This work was supported in part by the Natural Sciences and Engineering Research Council of Canada Discovery Grants DG-RGPIN-2022-03051 and DG-RGPIN-2023-04486. This research received funding from the European Research Council under the European Union's Horizon 2020 research and innovation program number 101002408 (MOS100PC). This work includes results based on observations obtained at the international Gemini Observatory, a program of NSF’s NOIRLab, which is managed by the Association of Universities for Research in Astronomy (AURA) under a cooperative agreement with the National Science Foundation on behalf of the Gemini Observatory partnership: the National Science Foundation (United States), National Research Council (Canada), Agencia Nacional de Investigaci\'{o}n y Desarrollo (Chile), Ministerio de Ciencia, Tecnolog\'{i}a e Innovaci\'{o}n (Argentina), Minist\'{e}rio da Ci\^{e}ncia, Tecnologia, Inova\c{c}\~{o}es e Comunica\c{c}\~{o}es (Brazil), and Korea Astronomy and Space Science Institute (Republic of Korea). This work has made use of data from the European Space Agency (ESA) mission {\it Gaia} (\url{https://www.cosmos.esa.int/gaia)}, processed by the {\it Gaia} Data Processing and Analysis Consortium (DPAC, \url{https://www.cosmos.esa.int/web/gaia/dpac/consortium}). Funding for the DPAC has been provided by national institutions, in particular the institutions participating in the {\it Gaia} Multilateral Agreement. Gemini spectra were processed using the DRAGONS package \citep{2023RNAAS...7..214L}.
\end{acknowledgments}

%\software{}

\facilities{Gaia DR3, Gemini-North (GMOS-N).}

%%%%%%%%%%%%%%%%%%%%%%%%%%%%%%%%%%%%%%%%%%%%%%%%%%%%%%%%%%%%%
%%%%%%%%%%%%%%%%%%%%%%%%%%%%%%%%%%%%%%%%%%%%%%%%%%%%%%%%%%%%%

\bibliographystyle{aasjournalv7}
\bibliography{main}

\end{document}